\documentclass{article}
\usepackage{spconf,amsmath,graphicx,subfigure,bbding}
\usepackage{booktabs,multirow,multicol}
\usepackage{cite}
\usepackage[colorlinks,urlcolor=black,citecolor=black, linkcolor=black]{hyperref}
\title{SUPER DENOISE NET: SPEECH SUPER RESOLUTION WITH NOISE CANCELLATION IN LOW SAMPLING RATE NOISY ENVIRONMENTS}
\name{Junkang Yang$^{1}$, Hongqing Liu$^{1,2}$,  Lu Gan$^{3}$, and Yi Zhou$^{1,2}$}
\address{$^{1}$ School of Communications and Information Engineering,\\ Chongqing University of Posts and Telecommunications (CQUPT), Chongqing, China\\
  $^{2}$ Intelligent Speech and Audio Research Lab (ISARL), CQUPT, Chongqing, China\\
  $^{3}$ Department of Electrical and Electronics Engineering, Brunel University, London, U.K.}
\begin{document}
%\ninept
%
\maketitle
\begin{abstract}
Speech super-resolution (SSR) aims to predict a high resolution (HR) speech signal from its low resolution (LR) corresponding part. Most neural SSR models focus on producing the final result in a noise-free environment by recovering the spectrogram of high-frequency part of the signal and concatenating it with the original low-frequency part. Although these methods achieve high accuracy, they become less effective when facing the real-world scenario, where unavoidable noise is present. To address this problem, we propose a Super Denoise Net (SDNet), a neural network for a joint task of super-resolution and noise reduction from a low sampling rate signal. To that end, we design gated convolution and lattice convolution blocks to enhance the repair capability and capture information in the time-frequency axis, respectively. The experiments show our method outperforms baseline speech denoising and SSR models on DNS 2020 no-reverb test set with higher objective and subjective scores. 
\end{abstract}
\section{Introduction}
\label{sec:intro}

Speech super-resolution (SSR) is the task to reconstructing the high-frequency part of the speech signal from its low-frequency part, which can be considered as bandwidth extension (BWE) in frequency domain. As one of the important tasks in the front-end of speech processing, SSR is widely applied in wireless communication, speech recognition \cite{haws2019cyclegan}, text-to-speech \cite{yoneyama2023nonparallel}, to name a few.

After decades of developments, a large number of SSR neural models have been proposed. Early methods based on deep neural network (DNN) \cite{li2015deep} show an acceptable performance by directly simulating the nonlinear relationship between the high frequency part and the low frequency part to improve the speech quality. The convolutional structures and Recurrent Neural Networks (RNN) have also been adopted in subsequent research \cite{kuleshov2017audio, rakotonirina2021self}, and these works extend the receptive field with less computational cost, addressing the lack of features captured by feed-forward convolutional models as well as the problem of vanishing gradients. In terms of feature dimensions, most models choose to operate in the raw waveform domain % \cite{kuleshov2017audio,su2021bandwidth,li2021real,lee2021nu,han2022nu} 
or in the frequency domain using short-time Fourier Transform (STFT) as input features. In recent works, generative models such as generative adversarial networks and diffusion-based approaches have achieved good results \cite{han2022nu,yoneyama2023nonparallel,shuai2023mdctgan,lemercier2023analysing}. %For example, [1] solved the problem of low-resolution speech bandwidth matching, [2] achieves high accuracy when parallel data from the target domain are unavailable, [3] uses the modified discrete cosine transform (MDCT) as input features. 

\begin{figure}%[htbp]
\centering
\subfigure[]
{
    \begin{minipage}[b]{.3\linewidth}
        \centering
        \includegraphics[scale=0.2]{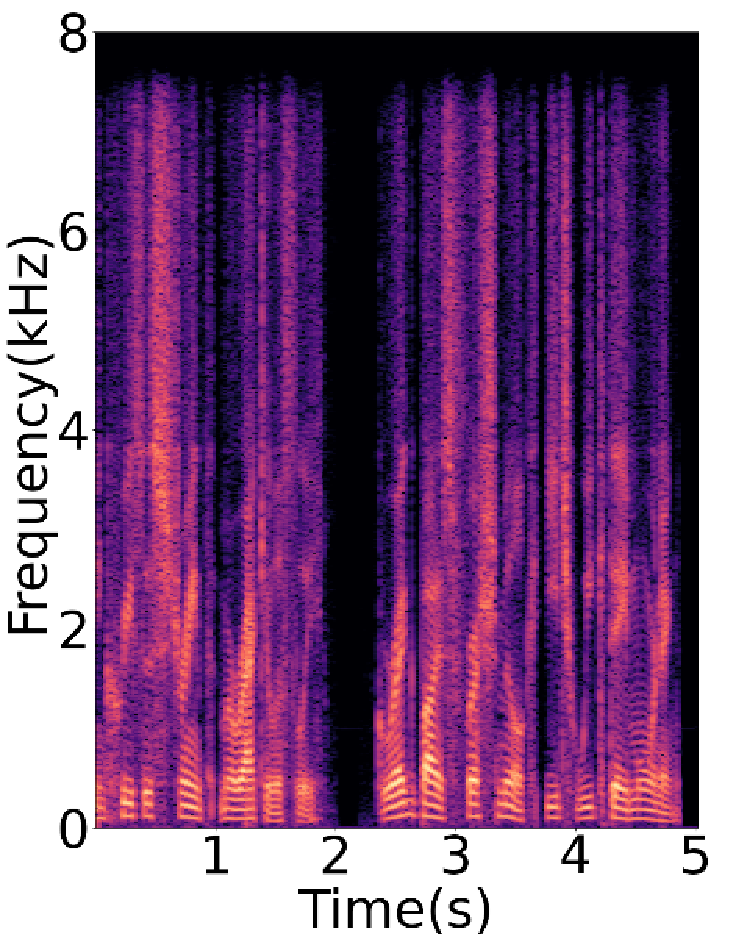}
    \end{minipage}
}
\subfigure[]
{
 	\begin{minipage}[b]{.3\linewidth}
        \centering
        \includegraphics[scale=0.2]{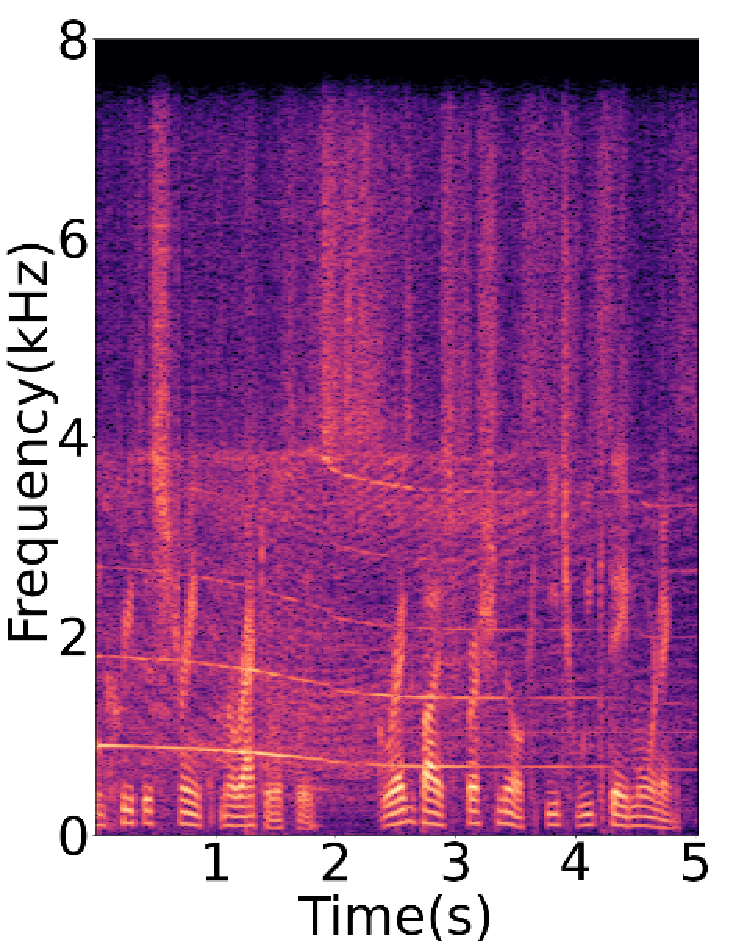}
    \end{minipage}
}
\subfigure[]
{
 	\begin{minipage}[b]{.3\linewidth}
        \centering
        \includegraphics[scale=0.2]{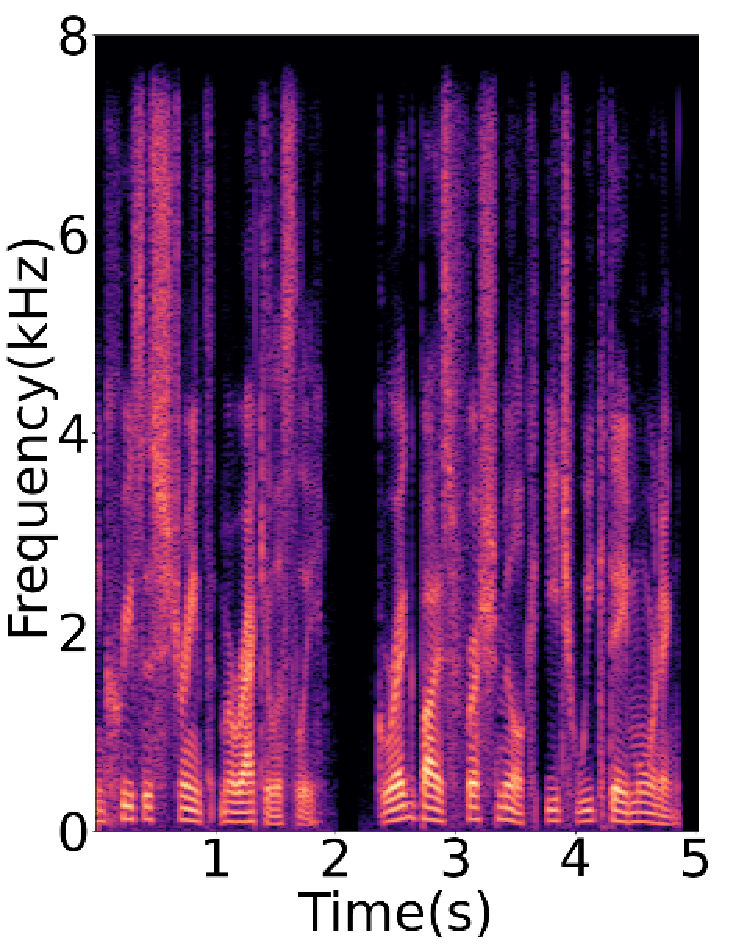}
    \end{minipage}
}
\caption{Spectrograms of reconstructed and original signals. (a) Reconstruction from the clean input. (b) Recovery from the noisy input. (c) Original high-resolution ground truth.}\label{fig1}
\end{figure}
Although SSR has achieved considerable progress, few studies have focused on the task of SSR in noisy environments, i.e., bandwidth extension along with the noise suppression. Existing frequency domain based work basically keeps the low-frequency part and predicts only the high-frequency part, and finally concatenates the two parts in series. In a noise-free environment, this concept can achieve good results, but when noise exists in the low-frequency part, it not only fails to remove the noise, but also produces the biased prediction of high-frequency part due to noise interference. This situation is illustrated in Figure \ref{fig1}, where the noisy high-resolution signal is generated due to the noise, see \ref{fig1}(b). Since most of the training data do not contain noise, the robustness of the model in noisy environments is bound to be greatly reduced. In \cite{seltzer2005robust}, the authors proposed an algorithm of robust BWE algorithm for speech with additive noise, transforming the narrowband spectral envelope into a wideband one in mel frequency cepstral coefficients (MFCC) domain. In \cite{seo2014maximum}, the authors introduced a multi-stage model, each of which focuses on noise reduction, corpus matching, and bandwidth extension, respectively. By predicting the intermediate representation at analyze stage and reconstructing the entire spectrogram with a GAN at synthesis stage, this concept \cite{liu2021voicefixer} can be used for general speech restoration, including SSR, denoising, dereverberation and inpainting. In \cite{chen2022deep}, authors combine UNet+AFiLM \cite{rakotonirina2021self} and an improved DTLN \cite{westhausen2020dual} to form a two-stage system. The authors in \cite{taher2023joint} simultaneously estimate the missing components and the noise distribution in degraded speech signal with a DNN.

\begin{figure}%[htbp]
\centering
\subfigure[]
{
    \begin{minipage}[b]{.45\linewidth}
        \centering
        \includegraphics[scale=0.3]{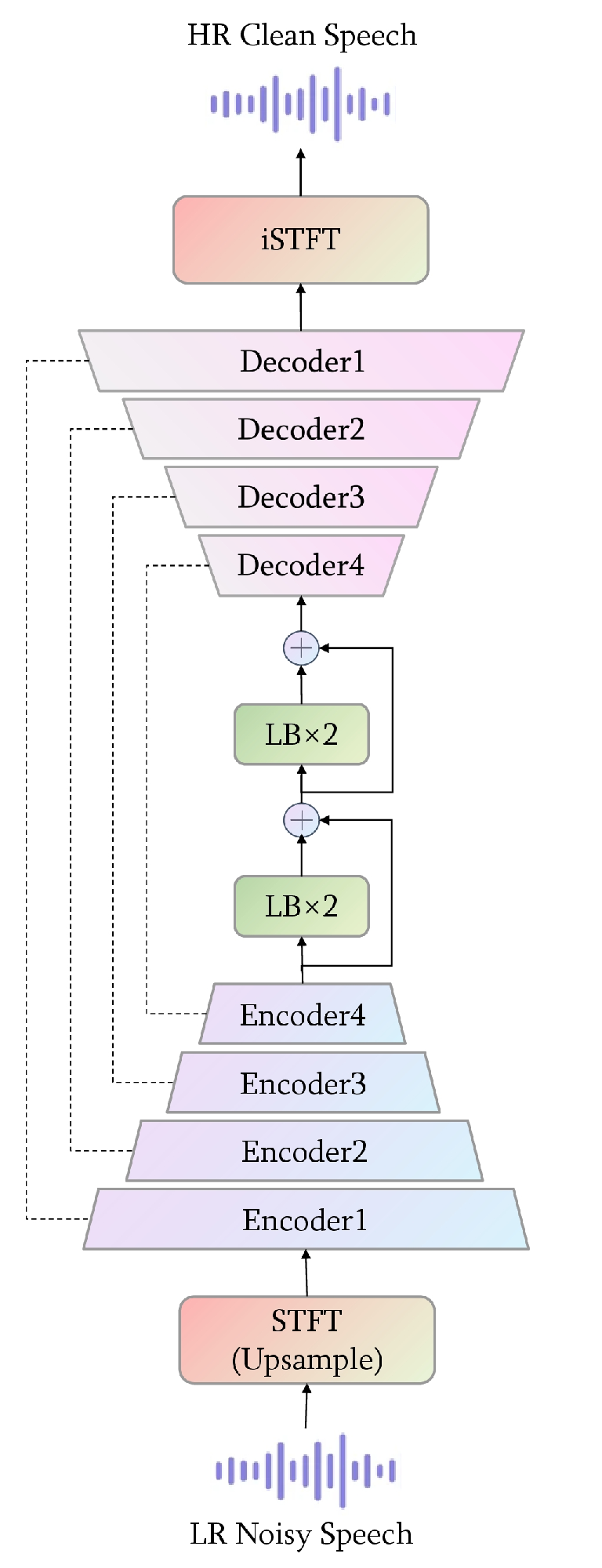}
    \end{minipage}
}
\subfigure[]
{
 	\begin{minipage}[b]{.45\linewidth}
        \centering
        \includegraphics[scale=0.3]{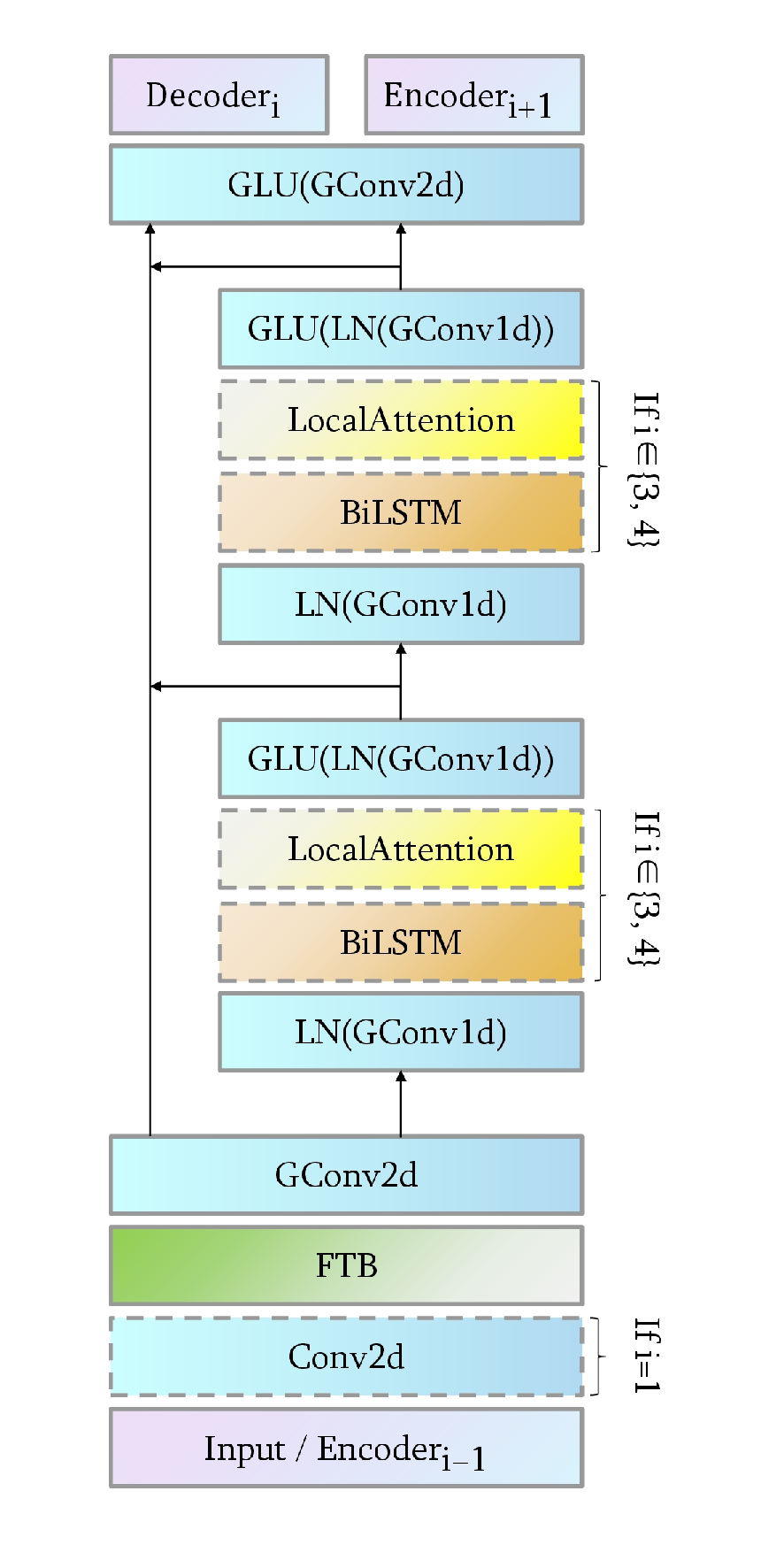}
    \end{minipage}
}

\caption{Architectures of (a) Network and (b) Encoder}\label{fig2}
\end{figure}

To improve the performance of SSR in noisy environments, we propose Super Denoise Net (SDNet), a neural network that removes noise while extending the bandwidth. Inspired by image restoration \cite{yu2019free,luo2022lattice}, we introduce gated convolution to enhance the generative capability of the network, and set lattice convolution blocks in the bottleneck layer to gain more information in the time-frequency domain. We train the model with large datasets containing noises. The experiments show our method substantially outperforms existing SSR models in evaluation metrics such as Perceptual Evaluation of Speech Quality (PESQ), Short-Time Objective Intelligibility (STOI), and DNSMOS, and even outperforms some models only focusing on noise reduction. Additionally, the ablation study demonstrates the impacts of our design on model performance.

\section{Methods}
\label{sec:methods}

\subsection{Problem Settings}
\label{ssec:problem}

In this paper, we address the problem of recovering HR clean speech from LR noisy speech. Given a HR clear speech $y \in R^{T}$, where $T$ is the length, after downsampling $s$ times and adding noise at the same sampling rate with the same length, the LR noisy speech $x_{noisy} \in R^{\frac{T}{s}}$ is generated. Our goal is to design a function $f$ that can efficiently predict $y$ from the observation $x_{noisy}$, i.e., $\hat{y}=f(x_{noisy})\approx{y}$. The same representation as above will be used in the formulas below.

\subsection{Network Architecture}
\label{ssec:net}

Our model uses a U-shaped structure similar to \cite{mandel2023aero}, containing encoders, decoders, and lattice convolution blocks (LBs) as bottleneck layers. The residual connections are set between encoders and decoders, also between the lattice convolution blocks. Model architecture is visualized in Figure \ref{fig2} (a). Due to space limit, we provide parameter settings, input size, and output size for each layer at our demo page\footnote{\url{https://sdnetdemo.github.io/}}.

\noindent
\textbf{Upsample.} The network operates in the spectral domain, and Short-time Fourier Transform (STFT) to the waveform input is required. Since the network predicts the full spectrogram, the signal needs to be upsampled at this stage to ensure that the input and output spectral are of the same scale. We follow the upsample method in \cite{mandel2023aero}, using a window length and hop length that is $\frac{1}{s}$ of those at the iSTFT stage while keeping the FFT bins constant. The window length, hop length, and FFT bins of iSTFT are 512, 512, and 64. We also move the complex part to the channel dimension at STFT stage.

\noindent
\textbf{Encoder and decoder.} As illustrated in Figure \ref{fig2}, there are 4 layers in encoder and decoder each. The input of encoder is the spectrogram of upsampled waveform and it will be reshaped at the first encoder layer with a 2D convolution. After that, a frequency transform block (FTB) \cite{yin2020phasen} is applied to capture the non-local correlations in spectrogram along the frequency axis. Unlike previous work, we utilize gated convolution (GConv) \cite{yu2019free} instead of ordinary convolution in the later structure, which enhances the model generation by learning a dynamic feature selection mechanism for each channel and each spatial location. Inside the encoder, there are two residual branches with two 1D gated convolutions at the beginning and end. In the middle, there are LSTM and temporal-based attention modules to capture long-distance relations. Followed by each encoder layer is a decoder layer that recovers the latent vectors equal to the size of spectrogram before passing the encoder. Its structure is the same as \cite{mandel2023aero}. It is worth noting that there is concatenated residual connection between each encoder and decoder layer, while between two residual branches within the encoder layer, it is a summation residual connection.

\begin{figure}%[htbp]
\centering
\includegraphics[scale=0.53]{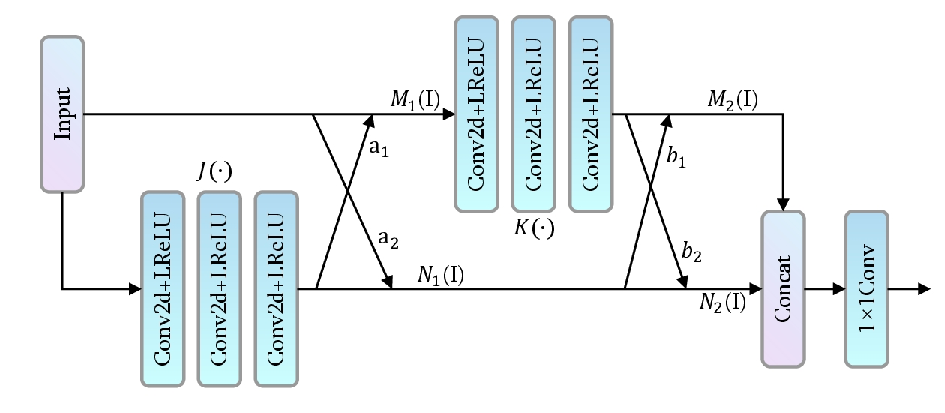}
\caption{Lattice convolution block (LB).}\label{fig3}
\end{figure}

\noindent
\textbf{Bottleneck layers.} The bottleneck layers include 4 lattice convolution blocks (LBs), which were first proposed in image restoration task \cite{luo2022lattice}. As shown in Figure \ref{fig3}, each LB includes paired butterfly-style structures. The input passes two branches that contain several convolution layers and LeakyReLU activation layer is followed for each layer. The two branches interact with each other through learnable combination coefficients. Specifically, given an input feature $I$, the first combination is 
\begin{equation}\label{eqn-1} 
  M_{1}(I) = I+a_{1}J(I),
\end{equation}
\begin{equation}\label{eqn-2} 
  N_{1}(I) = a_{2}I+J(I),
\end{equation}
where $J(\cdot)$ denotes to the implicit non-linear function of several layers shown in Figure \ref{fig3}. Similarly, the second combination is
\begin{equation}\label{eqn-3} 
  M_{2}(I) = b_{1}N_{1}+K(M_{1}(I)),
\end{equation}
\begin{equation}\label{eqn-4} 
  N_{2}(I) = N_{1}+b_{2}K(M_{1}(I)).
\end{equation}
Afterwards, the output of two branches are merged in channel dimension and then compressed through a 1\texttimes 1 convolution layer. The final output is 
\begin{equation}\label{eqn-5}
\begin{split}
  % O &= Conv(Concat(\\
  % & b_{1}(a_{2}I+J(I))+K(I+a_{1}J(I)), \\
  % & a_{2}I+J(I)+b_{2}K(I+a_{1}J(I)))).
  O &= Conv(Concat(M_{2}(I), N_{2}(I))).
\end{split}
\end{equation}
The combination coefficients are mainly determined in the following way. The mean and standard deviation in channel dimension are first obtained by global mean pooling in the upper branch and global standard deviation pooling in the lower branch. Then, those statistics in the two branches are passed through two fully connected layers, each followed by ReLU and Sigmoid activation functions, respectively. Finally, the outputs of the two branches are averaged to obtain the combined coefficients. The LB structure can dynamically adjust the dominance of different branches to fit different data through learnable combination coefficients, and its number of parameters is small. The combination of gated convolutions and lattice blocks could offer a blend of structured interpolation with adaptive, long-range context modeling, making it a potentially powerful tool for enhancing speech signals.

\subsection{Loss Function}
\label{ssec:loss}
We draw on the training methods of \cite{defossez2020real} and \cite{mandel2023aero}, where the model is trained using an adversarial approach. We also use a multi-scale STFT loss with FFT bins $\in \left\{512, 1024, 2048 \right\}$ and hop length $\in \left\{50, 120, 240 \right\}$ to jointly train the model. The window lengths are $\left\{240, 600, 1200 \right\}$. On the other hand, the multi-scale adversarial and feature losses %\cite{li2021real} 
in time domain is added in. The total loss can be expressed as
\begin{equation}\label{eqn-6} 
  \mathcal{L} = \mathcal{L}_{MSTFT} + \mathcal{L}_{adv} + \mathcal{L}_{f},
\end{equation}
where $\mathcal{L}_{MSTFT}$, $\mathcal{L}_{adv}$ and $\mathcal{L}_{f}$ are multi-scale STFT loss, adversarial loss, and feature loss.
\section{Experiments}
\label{sec:experiments}
\subsection{Dataset}
\label{ssec:dataset}
We use the dataset provided by Deep Noise Suppression (DNS) Challenge at ICASSP 2023 \cite{dubey2023icassp} to generate training data containing noise. We synthesize clean and noisy speech pairs with 500 hours by randomly mixing the speech and noise, and each sample lasts for 5 seconds. The SNR of all samples are at -5 - 20 dB and the sampling rate is 16 kHz. We then downsample all the noisy speech by a factor of $s=2$, i.e., turning the paired data into 8 kHz noisy speech and 16 kHz clean speech. For validation, we separate \%10 (50 hours) of total data as validation set. For testing, the no reverb test set of DNS Challenge 2020 \cite{reddy2020interspeech} is used and the noisy samples are also downsampled if neccessary.
\subsection{Baselines}
\label{ssec:baselines}
Our model is a multitasking model, so we selected baselines from SSR models, noise reduction models, and models that consider both tasks. For SSR models, we compare our model to WSRGlow \cite{zhang2021wsrglow} and NU-Wave 2 \cite{han2022nu}. For denoising models, our model is compared with DCCRN \cite{hu2020dccrn}, FullSubNet \cite{hao2021fullsubnet}, and DPT-FSNet \cite{dang2022dpt}. For multitasking models, the baselines are VoiceFixer \cite{liu2021voicefixer} and the network proposed in \cite{chen2022deep}. In our ablation studies, we retrain AERO \cite{mandel2023aero} with the training set mentioned in section \ref{ssec:dataset} as the baseline.
\subsection{Evaluation Metrics}
\label{ssec:evaluation}
\begin{table*}[htbp]
  \centering
  \caption{Performance evaluations, where $S$ means super-resolution only, $D$ indicates denoise only, and $SD$ means super-resolution and denoise.}
  \vspace{0.3cm}
    \begin{tabular}{c|c|c|c|ccc|c}
    \toprule
    Method & Task  & Sampling Rate of Source & Noise & PESQ  & STOI (\%)  & DNSMOS & MOS \\
    \midrule
    Noisy & —     & 16 kHz & \Checkmark     & 2.45     & 91.5     & 2.48     & — \\
    \midrule
    WSRGlow \cite{zhang2021wsrglow} & \multirow{4}[2]{*}{$S$} & \multirow{4}[2]{*}{8 kHz} & \XSolid     & 4.24     & 99.4     & 3.36     & — \\
    WSRGlow \cite{zhang2021wsrglow} &       &       & \Checkmark     & 2.44     & 91.0     & 2.47     & — \\
    NU-Wave 2 \cite{han2022nu} &       &       & \XSolid     & 4.22     & 99.4     & 3.20     & — \\
    NU-Wave 2 \cite{han2022nu} &       &       & \Checkmark     & 2.48     & 91.0     & 2.34     & — \\
    \midrule
    DCCRN \cite{hu2020dccrn} & \multirow{3}[2]{*}{$D$} & \multirow{3}[2]{*}{16 kHz} & \multirow{3}[2]{*}{\Checkmark} & 3.17     & 92.9     & —     & — \\
    FullSubNet \cite{hao2021fullsubnet} &       &       &       & 3.28     & 95.3     & —     & — \\
    DPT-FSNet \cite{dang2022dpt} &       &       &       & 3.28     & 95.3     & —     & — \\
    \midrule
    VoiceFixer \cite{liu2021voicefixer} & \multirow{3}[2]{*}{$SD$} & \multirow{3}[2]{*}{8 kHz} & \multirow{3}[2]{*}{\Checkmark} & 2.79     & 92.0     & 3.24     & 3.83 \\
    UNet + I-DTLN \cite{chen2022deep} &       &       &       & 2.78     & 91.3     & 2.93     & 3.01 \\
    \textbf{Proposed} &       &       &       & \textbf{3.52}     & \textbf{97.2}     & \textbf{3.36}     & \textbf{4.38} \\
    \bottomrule
    \end{tabular}%
  \label{tab:addlabel}%
\end{table*}%

\begin{table}[htbp]
  \centering
  \caption{Ablation studies on network achitechture.}
  \vspace{0.3cm}
    \begin{tabular}{l|cc}
    \toprule
    \multicolumn{1}{c|}{Mothod} & PESQ  & STOI (\%) \\
    \midrule
    \multicolumn{1}{c|}{AERO (retrained) \cite{mandel2023aero}} & 3.469     & 96.6 \\
                \qquad \qquad + GConv & 3.377     & 96.6 \\
                \qquad \qquad + LBs & 3.477     & 96.7 \\
                \qquad \qquad + Both & 3.522     & 97.2 \\
    \bottomrule
    \end{tabular}%
  \label{tab:addlabel}%
\end{table}%

To evaluate the quality of the generated speech, both objective and subjective evaluation metrics were used. The objective evaluation metrics used were PESQ \cite{recommendation2001perceptual}, STOI \cite{taal2011algorithm}, and DNSMOS \cite{reddy2021dnsmos}, where DNSMOS was shown to be highly correlated with human evaluation habits \cite{reddy2021dnsmos}. The subjective evaluation metrics is Mean Opinion Score (MOS) \cite{itu2003subjective}. We randomly selected 50 samples from the test set and asked 15 people to provide MOS values. For all evaluation metrics, the higher score represents the better listening experience.

\subsection{Results}
\label{ssec:results}
For all the models on denoise task only, we have referred to the results provided in \cite{li2023u}. As for other models, except the model in \cite{chen2022deep} and \cite{mandel2023aero}, we use the hyperparameters offered by the author to generate the target speech to evaluate the results. Since the authors did not provide the source code, we have re-implemented the method proposed in \cite{chen2022deep} to obtain the results.

Our experiments show that previous SSR models can achieve high PESQ scores and almost perfect STOI scores in noise-free environment, indicating that the generated speech is fully understandable. At the same time, the experiment also verified that when noise is present, their performance degrade sharply, and the generated speech may not even sound as good as the original noisy speech. In noisy environments, recent multitasking models have performed better on objective metrics, but based on the MOS scores provided by the participants, people still feel hard to know the information in speech. The results of our proposed model are superior to all baseline models in both subjective and objective evaluation indicators. Surprisingly, the proposed model even outperformed baseline models that only target denoising task. 

We also tested our model with real-world data, and the results showed that the noise in these speeches was effectively suppressed and the human voice was much clearer. Due to spcae limit, we provide the audio samples at our demo page. The demo contains 10 samples from the test set as well as some voice clips taken from early English interviews or speeches. 

\subsection{Ablation Studies}
\label{ssec:ablation}
In order to study the impact of network components on network performance, we conducted ablation studies by gradually adding each component to the original model. The experiments show that the original model actually achieves a decent result after training with a new dataset due to the approach of predicting all bands. However, when gated convolutions are introduced alone, the performance of network declines due to the absence of inverse components, which makes the first half of the network bloated. When LBs only is added, a small amount of improvements is produced, and when both components act together, the model achieves the best results.% However, the experiments demonstrated that the gated convolutions and LBs used produce a small amount of progress when introduced alone, and when they act together, the model achieves the best results.

\section{Conclusion \& Future Work}
\label{sec:typestyle}
In this paper, we proposed SDNet, a U-shaped encoder-decoder neural network to jointly handle SSR and noise suppression tasks in low sampling rate noisy environments. Our method predicts both high- and low-frequency parts of the signal and presents a strong generation capability. We have demonstrated through experiments that our model outperforms all baseline models in both objective and subjective evaluation metrics. The ablation studies have verified that our use of gated convolutions and LBs enhances the performance of the model. In our future work, we will focus on this joint task processing at higher resolutions, such as from 8 kHz to 24 kHz and/or from 16 kHz to 48 kHz. In terms of data, we will consider adding music and other personalized dataset.% which has different features compared with speech of human beings. % In addition, we will investigate the instability of loss values that occasionally occurs during model adversarial training.

% Below is an example of how to insert images. Delete the ``\vspace'' line,
% uncomment the preceding line ``\centerline...'' and replace ``imageX.ps''
% with a suitable PostScript file name.
% -------------------------------------------------------------------------
% \begin{figure}[htb]

% \begin{minipage}[b]{1.0\linewidth}
%   \centering
%   \centerline{\includegraphics[width=8.5cm]{image1}}
% %  \vspace{2.0cm}
%   \centerline{(a) Result 1}\medskip
% \end{minipage}
% %
% \begin{minipage}[b]{.48\linewidth}
%   \centering
%   \centerline{\includegraphics[width=4.0cm]{image3}}
% %  \vspace{1.5cm}
%   \centerline{(b) Results 3}\medskip
% \end{minipage}
% \hfill
% \begin{minipage}[b]{0.48\linewidth}
%   \centering
%   \centerline{\includegraphics[width=4.0cm]{image4}}
% %  \vspace{1.5cm}
%   \centerline{(c) Result 4}\medskip
% \end{minipage}
% %
% \caption{Example of placing a figure with experimental results.}
% \label{fig:res}
% %
% \end{figure}

%\vfill
%\pagebreak


\bibliographystyle{IEEEbib}
\bibliography{refs}

\end{document}